\documentclass[aps,prb,twocolumn,superscriptaddress]{revtex4-1}
\usepackage{amsfonts}
\usepackage{amsmath}
\usepackage{amssymb}
\usepackage{bm}
\usepackage{color}
\usepackage[usenames,dvipsnames,svgnames,table]{xcolor}
\usepackage{graphicx}
\usepackage{float}
\usepackage[varg]{txfonts}
\usepackage{wrapfig}
\usepackage{lmodern}
\usepackage{varwidth}
\usepackage{dcolumn}

\begin{document}

\title{Enhanced spin correlations in the Bose-Einstein condensate compound Sr$_3$Cr$_2$O$_8$}

\author{T.~Nomura}
\altaffiliation[Present address: ]{Institute for Solid State Physics, University of Tokyo, Kashiwa, Chiba 277-8581, Japan}
\affiliation{Dresden High Magnetic Field Laboratory (HLD-EMFL) and W\"urzburg-Dresden Cluster of Excellence ct.qmat, Helmholtz-Zentrum Dresden-Rossendorf, 01328 Dresden, Germany}

\author{Y.~Skourski}
\affiliation{Dresden High Magnetic Field Laboratory (HLD-EMFL) and W\"urzburg-Dresden Cluster of Excellence ct.qmat, Helmholtz-Zentrum Dresden-Rossendorf, 01328 Dresden, Germany}

\author{D.~L.~Quintero-Castro}
\altaffiliation[Present address: ]{Department of Mathematics and Natural Science, University of Stavanger, 4036 Stavanger, Norway}
\affiliation{Helmholtz-Zentrum Berlin f\"{u}r Materialien und
Energie, 14109 Berlin, Germany}

\author{A.A. Zvyagin}
\affiliation{Max-Planck-Institut f\"ur Physik komplexer Systeme, Noethnitzer Str., 38, D-01187, Dresden, Germany}
\affiliation{B.I.~Verkin Institute for Low Temperature Physics and 
Engineering of the National Academy of Sciences of Ukraine,
Nauky Ave., 47, Kharkiv, 61103, Ukraine}

\author{A. V.~Suslov}
\affiliation{National High Magnetic Field Laboratory, Tallahassee, Florida 32310, USA}

\author{D.~Gorbunov}
\affiliation{Dresden High Magnetic Field Laboratory (HLD-EMFL) and W\"urzburg-Dresden Cluster of Excellence ct.qmat, Helmholtz-Zentrum Dresden-Rossendorf, 01328 Dresden, Germany}

\author{S. Yasin}
\altaffiliation[Present address: ]{College of Engineering and Technology, American University of the Middle East, Kuwait}
\affiliation{Dresden High Magnetic Field Laboratory (HLD-EMFL) and W\"urzburg-Dresden Cluster of Excellence ct.qmat, Helmholtz-Zentrum Dresden-Rossendorf, 01328 Dresden, Germany}

\author{J. Wosnitza}
\affiliation{Dresden High Magnetic Field Laboratory (HLD-EMFL) and W\"urzburg-Dresden Cluster of Excellence ct.qmat, Helmholtz-Zentrum Dresden-Rossendorf, 01328 Dresden, Germany}

\author{K. Kindo}
\affiliation{Institute for Solid State Physics, University of Tokyo, Kashiwa, Chiba 277-8581, Japan}

\author{A.~T.~M.~N.~Islam}
\affiliation{Helmholtz-Zentrum Berlin f\"{u}r Materialien und
Energie, 14109 Berlin, Germany}

\author{B. Lake}
\affiliation{Helmholtz-Zentrum Berlin f\"{u}r Materialien und
Energie, 14109 Berlin, Germany}
\affiliation{Institut f\"{u}r Festk\"{o}rperphysik, Technische
Universit\"{a}t Berlin, 10623 Berlin, Germany}

\author{Y. Kohama}
\affiliation{Institute for Solid State Physics, University of Tokyo, Kashiwa, Chiba 277-8581, Japan}

\author{S.~Zherlitsyn}
\affiliation{Dresden High Magnetic Field Laboratory (HLD-EMFL) and W\"urzburg-Dresden Cluster of Excellence ct.qmat, Helmholtz-Zentrum Dresden-Rossendorf, 01328 Dresden, Germany}

\author{M. Jaime}
\altaffiliation[Present address: ] {Physikalisch-Technische Bundesanstalt, 38116 Braunschweig, Germany}
\affiliation{MPA-Maglab, Los Alamos National Laboratory, MS-E536, Los Alamos, New Mexico 87545, USA}

\begin{abstract}

Combined experimental and modeling studies of the magnetocaloric effect, ultrasound, and magnetostriction were performed on single-crystal samples of the spin-dimer system Sr$_3$Cr$_2$O$_8$ in large magnetic fields, to probe the spin-correlated regime in the proximity of the field-induced {\it XY}-type antiferromagnetic order also referred to as a Bose-Einstein condensate of magnons. The magnetocaloric effect, measured under adiabatic conditions, reveals details of the field-temperature ($H,T$) phase diagram, a dome characterized by critical magnetic fields $H_{c1}$ = 30.4 T,    $H_{c2}$ = 62 T, and a single maximum ordering temperature $T_{{\rm max}}(45~$T$)\simeq$~ 8 K. The sample temperature was observed to drop significantly as the magnetic field is increased, even for initial temperatures above $T_{{\rm max}}$, indicating a significant magnetic entropy associated to the field-induced closure of the spin gap. The ultrasound and magnetostriction experiments probe the coupling between the lattice degrees of freedom and the magnetism in Sr$_3$Cr$_2$O$_8$. Our experimental results are qualitatively reproduced by a minimalistic phenomenological model of the exchange-striction by which sound waves renormalize the effective exchange couplings.

\end{abstract}

\maketitle

\section{Introduction}

Spin-1/2 dimerized quantum magnets are paramagnetic systems that can exhibit a spin-singlet ground state and a spin gap $\Delta$ to the first spin-triplet excited state. Zeeman splitting of the spin triplet in an external field leads to closing of the gap and emergence of {\it XY}-type antiferromagnetic order. This ordered state exists between two field-induced quantum phase transitions at critical fields $H_{c1}$ (onset), and $H_{c2}$ (magnetization saturation), at temperatures $T$ $\ll \Delta/k_B$ with $k_B$ the Boltzman constant. The field range, $H_{c1} < H < H_{c2}$, is given by the dispersion of the spin triplet state and is, hence, strongly correlated to the system symmetry and sensitive to the presence of geometrical frustration. The physics of quantum magnets at critical fields when $T \rightarrow$ 0 can be mathematically described as Bose-Einstein condensation (BEC) of magnons\cite{2014_zapf} (also called magnonic BEC) and, as such, it is affected by the dimensionality of the space where the spin excitations exist. These characteristics make quantum magnets model systems to study the interplay between symmetry, dimensionality, and frustration and their impact on a variety of unique ground states.

One example is the dimer system Sr$_3$Cr$_2$O$_8$ with hexagonal bilayers of Cr$^{5+}$ ($S$ = 1/2) ions stacked along the $c$-axis and the intradimer exchange interaction $J_{0}$ = 5.55 meV. The interdimer interaction plays an important role in this material reducing the spin gap to 2$J_{0}$/3.\cite{2012_diana} It has been concluded from inelastic neutron scattering (INS) experiments that the excitations behave collectively like a strongly correlated gas even at elevated temperatures.\cite{2012_diana} A field-induced quantum phase transition was observed at the lower critical field $H_{c1}$ = 30.4 T, corresponding to the closing of the spin gap. The phase boundary close to $H_{c1}$ follows a critical-exponent law of the type $T_c \propto$ ($H-H_c$)$^\alpha$, with $\alpha$ = 2/3, the expected value for BEC in three dimensions (3D). The upper critical field $H_{c2}$ = 62 T was determined as the field at which the magnetization $M(H)$ saturates.\cite{2009_aczel} A 3D canted-{\it XY} AFM order exists in a dome-like structure with a maximum ordering temperature of $T_{{\rm max}} \simeq$ 8 K at 45 T. \cite{2009_aczel} 

A Jahn-Teller transition at $T_{JT}$ = 285 K was found to lift the degeneracy of the singly occupied $e$ orbitals of Cr$^{5+}$ (partially removing the concomitant geometrical frustration) and lowering the crystal symmetry from hexagonal ($R \overline{3} m$) to monoclinic ($C2/c$),\cite{2008_chapon,2010_wang} leading simultaneously to three crystallographic domains at low temperatures and a weakly interacting gas of triplet excitations. On the other hand, recent ultrasound and magnetization experiments performed in pulsed magnetic fields have suggested a magnonic liquid state above the dome,\cite{2016_wang,2016b_wang} a smoking gun for strong correlations persisting to cryogenic temperatures. These seemingly contradictory results call for a quantitative assessment of adiabatic magnetization effects that, through sample-temperature variations in the environment of rapidly changing pulsed magnetic fields, can teach us more about the microscopic mechanisms and their correlations. Also in need of attention is the spin-lattice coupling in Sr$_3$Cr$_2$O$_8$, often neglected in quantum magnets. 

In this work, we focus on the properties of the field-induced AFM dome, with special emphasis on spin excitations above the dome and spin-lattice coupling with probes sensitive to both microscopic and macroscopic mechanisms. We complement the experimental results with a phenomenological model for the exchange striction that reproduces well the magnetization, magnetic susceptibility, and sound propagation properties as function of temperature and magnetic fields. Our results on Sr$_3$Cr$_2$O$_8$ bring light onto aspects of quantum magnets that are not well understood and will impact other strongly correlated systems.

\section{Experimental Methods}

\begin{figure}
\centering
\includegraphics[width=0.98\linewidth]{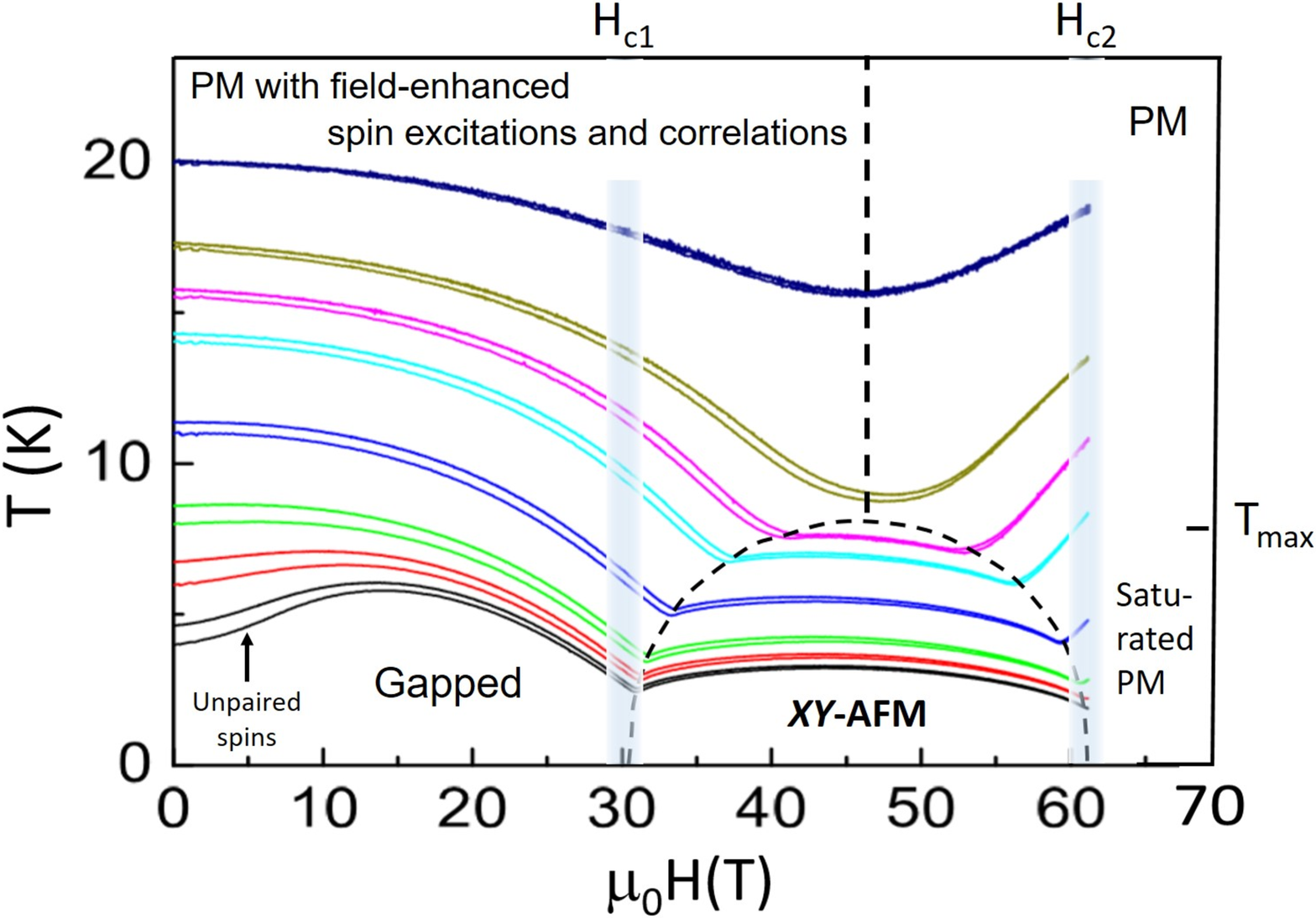}
\caption{Adiabatic magnetocaloric-effect data obtained for Sr$_3$Cr$_2$O$_8$ in a 36 ms short-pulse magnet, with the magnetic field applied along the crystallographic $c$-axis. Individual curves were obtained for various initial temperatures between 4 and 20 K. The lowest temperature curves show initial heating, possibly caused by the presence of a small number of paramagnetic impurities.\cite{2009_aczel} The sample temperature for all curves shows an overall cooling effect, which can be attributed to the combined effect of enhanced excitations and correlations as the spin energy-gap is suppressed by the increasing magnetic field.}
\label{fig1}
\end{figure}

High-quality single-crystal samples of Sr$_3$Cr$_2$O$_8$ were grown using the floating-zone technique. \cite{2010_islam} Spin-lattice effects in this material were studied by means of ultrasound (sound velocity) and dilatometry (magnetostriction and thermal expansion) techniques, complemented by magnetocaloric-effect experiments. A high-resolution phase-sensitive pulse-echo technique was employed in the ultrasound experiments.\cite{Zherlitsyn14} The longitudinal, $c_{11}$ ({\bf k}$\|${\bf u}$\|a$), $c_{33}$ ({\bf k}$\|${\bf u}$\|c$) and transverse $c_{44}$ ({\bf k}$\|a$, {\bf u}$\|c$), ($c_{11}-c_{12}$)/2 ({\bf k}$\|a$, {\bf u}$\|b$, $a \perp b$) acoustic modes were investigated in the ultrasound experiments. Here, {\bf k} and {\bf u} are the wave vector and polarization of the acoustic wave, respectively, and the crystallographic directions are given for the hexagonal crystal. The elastic moduli $c_{ij}$~=~$\rho v^{2}$ are proportional to the square of the velocities of the corresponding acoustic mode, $v$. Here $\rho$ is the mass density. Resonant LiNbO$_{3}$ and wide-band PVDF-film (polyvinylidene-fluoride film) transducers glued to the samples were used for sound generation and detection. The typical ultrasound frequencies were between 30 and 100 MHz.

The magnetostriction and thermal-expansion experiments were performed using an optical fiber Bragg (FBG) grating technique.\cite{2010_ramsy, 2017_jaime} Calibrated RuO$_{2}$ resistors, directly attached to the sample, were employed for the thermometry in the ultrasound and magnetostriction experiments. Continuous ({\it dc}) magnetic field experiments up to 45 T were performed at the NHMFL, USA.
 
The adiabatic magnetocaloric-effect (MCE) was measured using an Au-Ge thermometer sputtered directly on the sample.\cite{13RSI_Kihara,14PRB_Kohama} The resistance of the thermometer was measured by a standard {\it ac} four-probe method by means of a digital lock-in technique at a frequency of 50~kHz. The pulsed-field MCE experiments were performed at the ISSP, University of Tokyo, Japan.

\begin{figure}
\centering
\includegraphics[width=0.75\linewidth]{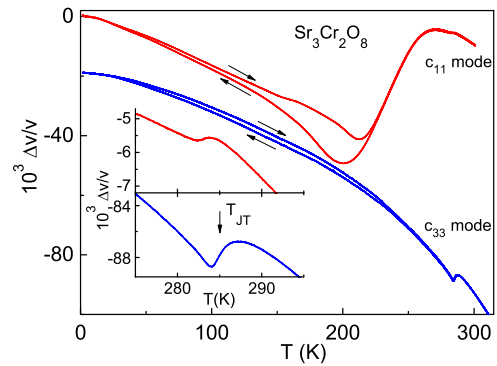}
\caption{Relative change of the sound velocity $\Delta v/v$ vs temperature for the longitudinal  $c_{11}$ and $c_{33}$ acoustic modes in Sr$_3$Cr$_2$O$_8$. The inset shows the vicinity of the Jahn-Teller phase transition. The ultrasound frequency was  60~MHz for the $c_{11}$ mode (red line) and 47 MHz for the $c_{33}$ mode (blue line).}
\label{fig2}
\end{figure}

\begin{figure}
\centering
\includegraphics[width=0.72\linewidth]{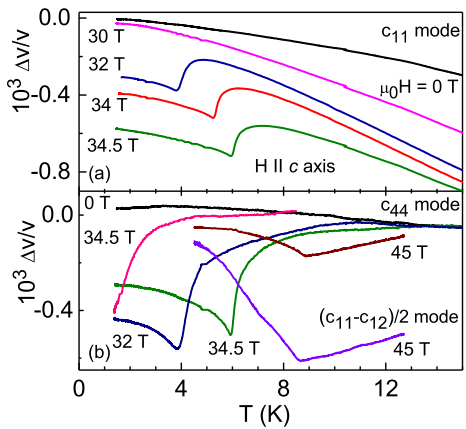}
\caption{Relative change of the sound velocity $\Delta v/v$ vs temperature  for (a) the longitudinal  $c_{11}$ mode, and (b) the transverse $c_{44}$, and ($c_{11}$-$c_{12}$)/2  acoustic modes in Sr$_3$Cr$_2$O$_8$  at selected magnetic fields.}
\label{fig3}
\end{figure}

\section{Results and Discussion}

\subsection{Temperature-field phase diagram from adiabatic magnetocaloric effect}

Figure  \ref{fig1} shows the temperature of a 21.4~mg sample mounted with the crystallographic $c$-axis along the applied magnetic field when the field is pulsed to 60~T in a capacitor bank-driven pulsed magnet. The sample in this experiment is adiabatically magnetized and, hence, its temperature does not remain constant. This phenomenon, known as magnetocaloric effect, is governed by the Maxwell relation ($\partial T/\partial H)_S = - T/C_H (\partial M/\partial T)_H$, where $S$ is the entropy, $C_H$ is the specific heat measured at constant field, $M$ is the magnetization, $T$ is the temperature, and $H$ is the magnetic field. This equation predicts that a negative slope in the sample $M(T)$ curve leads to a sample temperature increase while adiabatically magnetized, and the other way around when $\partial M/\partial T > 0$. The magnitude of the magnetocaloric effect is, additionally, larger at low temperatures when $C_H (T) \rightarrow 0$. With these considerations in mind we note that our data in Fig. \ref{fig1} are in excellent qualitative agreement with the known magnetic susceptibility $\chi(T)$.\cite{2009_aczel} We also note that the {\it XY}-AFM dome is evident, as phase boundaries at $H_c(T)$ are identified as dips in the $T(H)$ curves. We also note a significant drop in sample temperature above the dome,  i.e. $T > T_{max}$, which signals magnetic entropy accumulation. Similar entropy accumulation was observed in the quantum magnet Cu(pyz)(gly)ClO$_4$ \cite{2016_brambleby} where $C_p(T,H)$ data is available to well above $T_{{\rm max}}$, and discussed in the framework of various Schottky-type anomalies in the thermal properties caused by Zeeman-split spin-triplet levels in the presence of a magnetic field.\cite{2014_zapf} A magnonic liquid state was recently proposed to address the physics of quantum magnets above the dome\cite{2016_wang, 2016b_wang} and, hence, we take in the following sections a closer look at the spin-lattice coupling in Sr$_3$Cr$_2$O$_8$ to find out if unusual mechanisms pointing to a spin-liquid type of state are at play. We do so by cutting through the ($T,H$) diagram in Fig. \ref{fig1} under $T$ = const. and $H$ = const. conditions in continuous magnetic fields to 45 T.

\subsection{Temperature dependent sound velocity in zero field}

Figure \ref{fig2} shows the temperature dependence of the sound velocity of the longitudinal $c_{11}$ and $c_{33}$ modes below 310 K in zero applied magnetic field. Both acoustic modes exhibit anomalies in the sound velocity of the order of 10$^{-3}$ at the Jahn-Teller phase transition $T_{JT} \sim$ 285 K. When lowering the temperature, the $c_{33}$ mode experiences a characteristic hardening due to the anharmonicity of the ionic potential,\cite{1970_varshni} though the $c_{11}$ mode manifests a more complex behavior. A rather large softening ($\approx$ 4 $\%$ in $\Delta v/v$) appears slightly below $T_{JT}$ down to 200 K. In addition, a large hysteresis is observed between 50 and 220 K. No further anomalies have been detected for the longitudinal acoustic modes down to the lowest measured temperature of 1.5 K in zero magnetic field, consistent with expectations.

\begin{figure}
\centering
\includegraphics[width=0.7\linewidth]{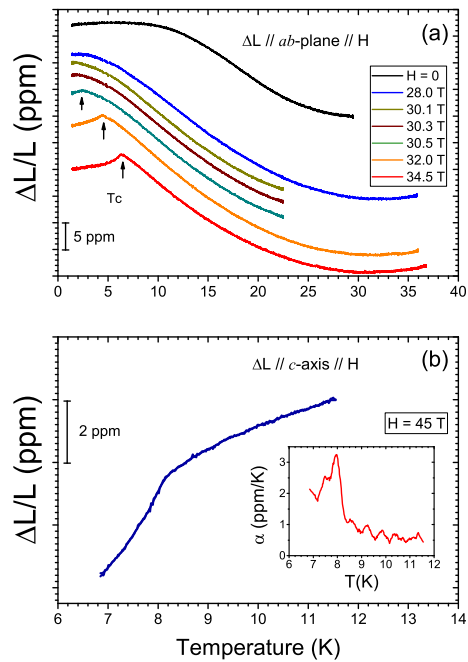}
\caption{(a) Thermal expansion measured in the $ab$-plane at various DC magnetic fields up to 34.5 T. (b) Thermal expansion measured along the $c$-axis direction at 45 T. Inset: Coefficient of thermal expansion vs. temperature  $\alpha$(T) computed at 45 T.}
\label{fig4}
\end{figure}

\subsection{Sound velocity and dilatometry in static magnetic fields}

The temperature dependencies of the sound velocity for various acoustic modes are shown in Fig. \ref{fig3} at selected magnetic fields applied along the [001] direction. For $H > H_{c1}$ clear anomalies are observed by entering the ordered phase. Remarkably, while the longitudinal $c_{11}$ mode shows only a well-localized step-like anomaly at the transition temperature (see Fig. \ref{fig3}(a)), the softening starts above 12 K for the transverse acoustic modes $c_{44}$ and ($c_{11}$-$c_{12}$)/2 (see Fig. \ref{fig3}(b)). This fact hints at the higher sensitivity of the transverse acoustic modes to enhanced spin-spin correlations existing in Sr$_3$Cr$_2$O$_8$ above the dome. Indeed, the $c_{44}$ curves measured at 32 T (dark blue) and 34.5 T (green) bear a striking resemblance to lambda-type  anomalies (just inverted), typical of second-order phase transitions where temperature-driven fluctuations above and below the ordered state are known to be important.

To further explore the role of the spin-strain interactions in the strongly correlated state we have performed longitudinal dilatometry experiments in Sr$_3$Cr$_2$O$_8$ in constant applied magnetic fields. Figure \ref{fig4}(a) shows the longitudinal thermal-expansion of the sample measured at various magnetic fields applied in the $ab$-plane. In zero field the smooth black curve shows no anomalies below 30 K, with an inflection point at a temperature consistent with the opening of the spin gap.\cite{2009_aczel} A clear anomaly is observed at the phase transition into the low-temperature ordered state for finite fields. Note, the lattice effect is only of the order of 10$^{-6}$. Hence, the sound velocity changes (see Fig. \ref{fig3}) cannot be explained alone by the crystal-lattice elongation/contraction in the proximity to the phase transition. The longitudinal thermal expansion along the crystallographic $c$-axis measured at the top of the dome in a 45 T fields is displayed in Fig. \ref{fig4}(b). Again, the observed shape is consistent with a lambda-type transition for the coefficient of thermal expansion (see Fig. \ref{fig4}, inset), although the scattering in the data is too high to rule out other possibilities.

\begin{figure}
\centering
\includegraphics[width=0.7\linewidth]{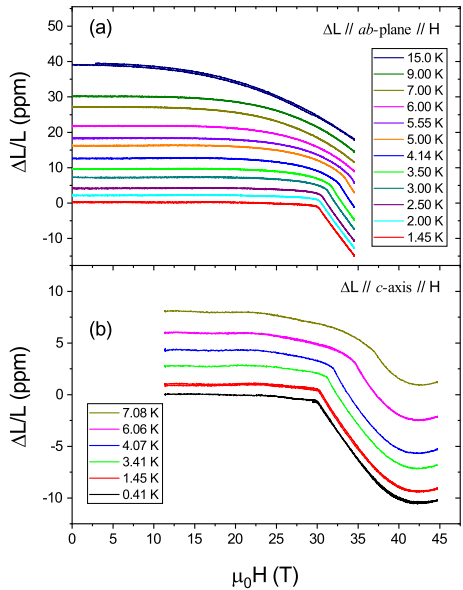}
\caption{Magnetic-field dependence of the axial strain measured in continuous magnetic fields (a) along the $ab$-plan direction and (b) along the $c$-axis direction at various temperatures. The small oscillations in the paramagnetic sate $H <H_{c1}$ are an artifact originated in Faraday rotation of light in the optical fiber caused by the large 45 T magnet fringe field. \cite{2017_jaime}}
\label{fig5}
\end{figure}

\begin{figure}
\centering
\includegraphics[width=0.7\linewidth]{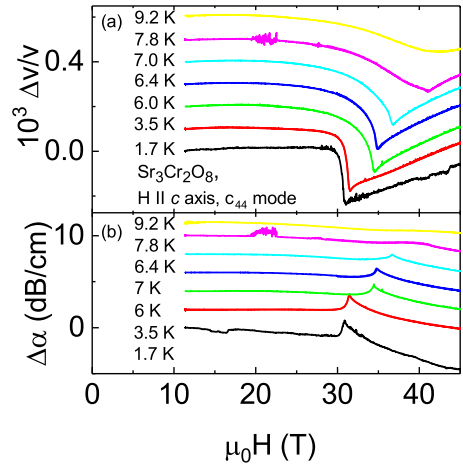}
\caption{Relative change of (a) the sound velocity and (b) sound attenuation of the $c_{44}$ mode versus magnetic field ($H\|c$) measured at selected temperatures in Sr$_3$Cr$_2$O$_8$.}
\label{fig6}
\end{figure}

\begin{figure}
\centering
\includegraphics[width=0.7\linewidth]{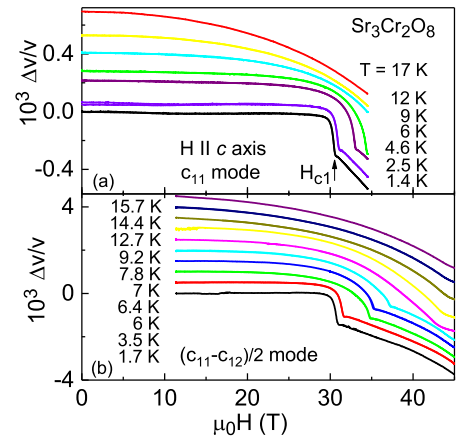}
\caption{Relative change of the sound velocity of (a) the longitudinal $c_{11}$ and (b) transverse ($c_{11}$-$c_{12}$)/2 mode versus magnetic field measured at selected temperatures in Sr$_3$Cr$_2$O$_8$.}
\label{fig7}
\end{figure}

\subsection{Sound velocity, attenuation, and dilatometry in sweeping magnetic field at constant temperature}

The longitudinal magnetostriction, $\Delta L/L$, measured for $H\| ab$-plane and $H\|c$-axis (Fig.~\ref{fig5}) shows anomalies at $H_{c1}$ and contraction of the respective lattice parameters that resemble the shrinkage in the order state displayed in the thermal expansion in Fig. \ref{fig4}. A significant contraction of the lattice with magnetic fields is evident at temperatures $T > T_{max}$, following a trend that resembles the MCE displayed in Fig. \ref{fig1}. Closing of the spin gap has, as primary consequence, the emergence of magnetic excitations. Since the exchange interaction among Cr$^{2+}$ atoms in Sr$_3$Cr$_2$O$_8$ is AFM, the lattice contracts with increasing field due to AFM fluctuations in the proximity of AFM ordering.

\begin{figure}
\centering
\includegraphics[scale=0.3]{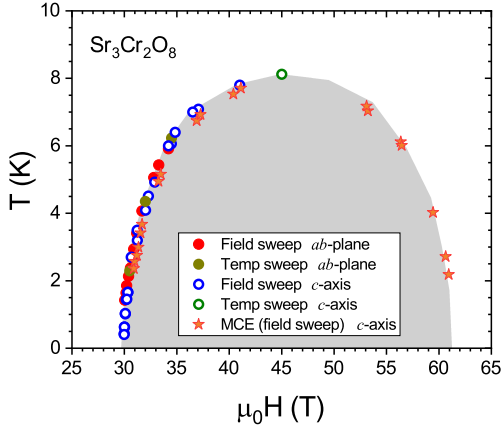}
\caption{($H,T$) phase diagram from our dilatometry and sound-velocity data for {\it dc} magnetic fields along the $ab$-plane and along the $c$-axis, and magnetocaloric effect data in pulsed magnetic fields. While excess entropy seems to have an important effect at temperatures $T > T_{{\rm max}}$, our diagram largely agrees with that of Aczel et al., \cite{2009_aczel} obtained on samples from a different source.
}
\label{fig8}
\end{figure}

Figures \ref{fig6} and \ref{fig7} exhibit magnetic field dependencies of the acoustic properties for the longitudinal $c_{11}$ and transverse $c_{44}$ and $(c_{11}-c_{12})/2$ modes from Fig. \ref{fig3} measured in static magnetic fields in two different magnets to 35 T and to 45 T at various temperatures. At $H_{c1}$, the transverse $c_{44}$ mode reveals a sharp minimum in the sound velocity (Fig. \ref{fig6}(a)) and a maximum in the sound attenuation (Fig. \ref{fig6}(b)), though the  longitudinal $c_{11}$ and transverse ($c_{11}$-$c_{12})/2$ modes show a step-like anomaly in the sound velocity (Fig. \ref{fig7}). No hysteresis is observed at the phase transition. The total sound-velocity change at $H_{c1}$ is of the order of $10^{-4}$, comparable for all studied acoustic modes. Transition temperatures measured at constant magnetic field (temperature sweep), critical fields obtained at constant temperatures (field sweep) in {\it dc} magnetic fields for magnetic fields oriented parallel to  the crystrallographic $c$-axis and the $ab$-plane are ploted in Fig. \ref{fig8}. The resulting (H,T) phase diagram agrees well with earlier pulsed field data by Aczel et al.\cite{2009_aczel}

\subsection{Modeling}

Two reasons can cause the renormalization of the sound velocity and attenuation in magnetic systems. The crystalline electric field of nonmagnetic ions (ligands), surrounding the magnetic ones, is affected by sound waves shifting in turn the positions of ligands. This way, the crystalline electric field together with the spin-orbit coupling, yields changes in the single-ion magnetic anisotropy of the magnetic ions. This effect has a relativistic origin, and the interaction between the sound waves and the magnetic material properties is present at any temperature lower than the characteristic energy of the single-ion magnetic anisotropy. This is not the case, however, for Sr$_3$Cr$_2$O$_8$ because the single-ion anisotropy does not occur in spin-1/2 systems. The second mechanism by which the sound propagation characteristics are affected by magnetism is the exchange striction. Sound waves change the positions of magnetic and/or nonmagnetic ions involved in the indirect exchange coupling (superexchange~\cite{Goodenough_1963}) and, in this way,  renormalize the effective exchange coupling between magnetic ions. To describe our experimental findings for the magnetic and sound characteristics in Sr$_3$Cr$_2$O$_8$ we model the exchange-striction mechanism that affects inter-spin exchange interactions and determine the magnetic phase transitions in the system. 

According to Tachiki~\cite{Tachiki_1974} the exchange-striction coupling in magnetic systems yields a renormalization of the velocity of the sound waves proportional to spin-spin correlation functions. Those correlation functions can be approximated by a combination of the magnetization and the magnetic susceptibility of the system. Good agreement between experiments and theory, even if only the homogeneous part of the magnetic susceptibility is taken into account, was achieved for many systems including magnetically ordered \cite{Andreev_2017}, low-dimensional \cite{Sytcheva_2009}, and spin-frustrated ones \cite{Zherlitsyn_2015}. Similar good agreement with experiment was also obtained in the analysis of the magneto-acoustic characteristics for a dimerized spin-1/2 system \cite{Chiatti_2008}. 

The renormalization of the sound velocity, $v$, due to the exchange-striction coupling in the general case can be written as 
\begin{eqnarray}
&&\frac {\Delta v}{v} \approx  - \frac {v}{\rho V \omega^2\mu^4} \times  \left[|g(0)|^2A + h(0)\mu^2B\right]
\end{eqnarray}
with 
\begin{eqnarray}
&& A = (2M^2\chi + k_BT\chi^2), B =  (M^2 +k_BT\chi)  \nonumber
\end{eqnarray}
where $V$ is the volume of the crystal, $\mu$  the effective magneton per magnetic ion, and $M$ the magnetization. The magneto-elastic coefficients are

\begin{eqnarray}
&&h(q) =\sum_j e^{-i{\bf q}{\bf R}_{ji}}\left[ 1-\cos ({\bf k}{\bf R}_{ji})\right] ({\bf u}_{\bf k}\cdot {\bf u}_{-{\bf k}})(\frac {\partial^2 J_{ij}^{\beta,\beta '}}{\partial {\bf R}_i \partial {\bf R}_j}), \ \nonumber \\ 
&&g(q) = \sum_j e^{i{\bf q}{\bf R}_{ji}}\left( e^{i{\bf k}{\bf R}_{ji}} -1\right) {\bf u}_{\bf k} \frac {\partial J_{ij}^{\beta,\beta '}}{\partial {\bf R}_i} 
\end{eqnarray}
(taken at ${\bf q}=0$), where ${\bf R}_{ji} = {\bf R}_j -{\bf R}_i$, ${\bf R}_j$ is the position vector of the $j$-th site of the magnetic ion, and $J_{ij}^{\beta,\beta '}$ ($\beta,\beta ' = x,y,z$) are the exchange couplings between magnetic ions on the $i$-th and $j$-th site. Similar results can be obtained for the sound attenuation,
\begin{eqnarray}
&&\Delta \alpha \approx  \frac {\gamma }{\rho V v \mu^4(\omega^2 +\gamma^2)} \times  \left[|g(0)|^2A + h(0)\mu^2B\right] \
\end{eqnarray}
where $\gamma$ is the effective relaxation rate. In our calculations, the magnetic field is directed along {the} $c$ axis of the crystal. 

\begin{figure}
\centering
\includegraphics[width=0.72\linewidth]{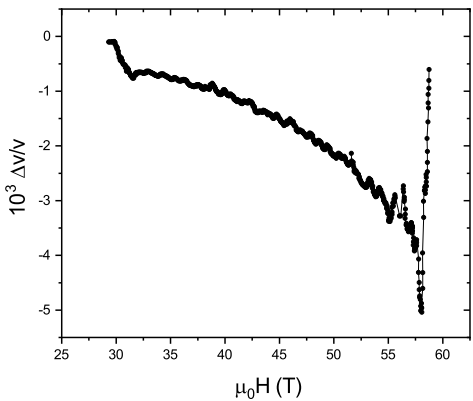}
\caption{Calculated magnetic field dependence for the relative changes in the sound velocity related to the shear mode $c_{44}$ for Sr$_3$Cr$_2$O$_8$ at $T=4.2$~K, using the known magnetization vs field. These results reproduce well the experiments~\cite{2016_wang,2016b_wang}. } 
\label{fig9}
\end{figure}

To check the applicability of the exchange-striction model for the description of the magneto-acoustic experiments in Sr$_3$Cr$_2$O$_8$, we use the known magnetization vs field data by Wang et al.~\cite{2016_wang,2016b_wang}. Figure~\ref{fig9} shows that the formula used to connect $\Delta v/v$ with  $M(H,T)$, and $\chi(H,T)$ works well for $T=4.2$~K, and the features associated to the critical fields at $H_{c1}  \approx 30.5$~T and $H_{c2} \approx 62$~T are evident.

In order to calculate the magnetization and magnetic susceptibility versus external magnetic field and temperature, we use a phenomenological fermionic model which consists of two branches of fermion excitations (see, Zvyagin et al.~\cite{Zvyagin_2018} for the dimerized spin chain), 
\begin{eqnarray} 
&&\varepsilon_{k}^{(1,2)} =\mu H \pm \frac{1}{2} [(H_{c2}+H_{c1})^2+ \nonumber \\
&&+ (H_{c2}-H_{c1})^2+(2(H_{c2}^2-H_{c1}^2)\cos(k)]^{1/2} \ 
\end{eqnarray} 

Two branches are used to represent the singlet pair. As pointed out by Giamarchi et al.\cite{Giamarchi_2008}, hard-core bosons behave similar to fermions in the spin-dimer systems. The first branch is gapped for any value of $k$. The second branch is gapped for $H < H_{c1}$ (corresponding to the Fermi point\cite{Fermi} at $H=H_{c1}$ for $k=\pi$) and for $H> H_{c2}$ (corresponding to the spin-polarized phase with the Fermi point at $H=H_{c2}$ for $k=0$). For $H_{c1} < H < H_{c2}$ the second branch is gapless. The behavior of the second branch in a magnetic field is reminiscent of the one observed in Sr$_3$Cr$_2$O$_8$ in neutron scattering experiments. \cite{Castro_2010,2012_diana}. The van Hove singularities at the critical fields $H=H_{c1,2}$ define quantum phase transitions between the gapped singlet, gapless, and gapped spin-polarized phases take place. In the framework of this model the magnetization per spin is 
\begin{equation}
M = \frac{\mu}{2N} \sum_{n=1}^2\sum_k \tanh \left[\frac{\varepsilon_k^{(n)}}{2k_BT}\right] \  
\end{equation}
and the magnetic susceptibility is 
\begin{eqnarray}
&&\chi = \frac {1}{2N}\sum_{n=1}^2 \sum_k \biggl( \frac {\partial^2 \varepsilon_k^{(n)}}{\partial H^2} 
\tanh \left[\frac {\varepsilon_{k}^{(n)}}{2k_BT} \right] + \nonumber \\
&&+ \left[\frac {\partial \varepsilon_k^{(n)}}{\partial H}\right]^2 \frac {1}{2k_BT\cosh^2 (\varepsilon_k^{(n)}/2k_BT)}  \biggr) \ 
\end{eqnarray}

Then, the three-dimensional interaction between chains, which produces the magnetic ordering in the spin system, is taken into account in the random phase approximation which yields  $\chi(H_{c},T_c) = 1/ZJ_{eff}$ for the critical temperatures and critical fields, where $Z$ is the coordination number, and $J_{eff}$ is the effective coupling between chains. The model is, in fact, the cluster version of the mean-field theory. It is known \cite{Castro_2010} that in Sr$_3$Cr$_2$O$_8$ spin dimers, while oriented along the $c$-axis of the crystal, do not interact as a quasi-one-dimensional system. Indeed, the interaction between spin dimers is more complicated, with dimers arranged in a hexagonal lattice. To justify somehow the one-dimensional cluster approach one can take into account that the effective coupling between spins of dimers along the $c$-axis is frustrated \cite{Castro_2010}, thus an effective coupling can replace real interactions between spins of dimers along $c$-axis.  Also, we note that for the low-dimensional spin system the quasi-one-dimensional approach better describes the magneto-acoustic characteristics than the quasi-two-dimensional one \cite{Sytcheva_2009}. We point out that our model yields a better description of the spin system of Sr$_3$Cr$_2$O$_8$ than the cluster mean-field-theory, based on mean-field-coupled spin dimers. For example, the dimer version of the mean-field \cite{Tachiki_1970} does not produce the gapless phase between $H_{c1}$ and $H_{c2}$ (in fact, all states in that simplified theory are related only to levels, not to band/extended states). Also, following Nikuni et al.~\cite{Nikuni_2000} we plot the temperature dependence of the susceptibility for several values of the magnetic field above $H_{c1}$.

\begin{figure}
\centering
\includegraphics[width=0.75\linewidth]{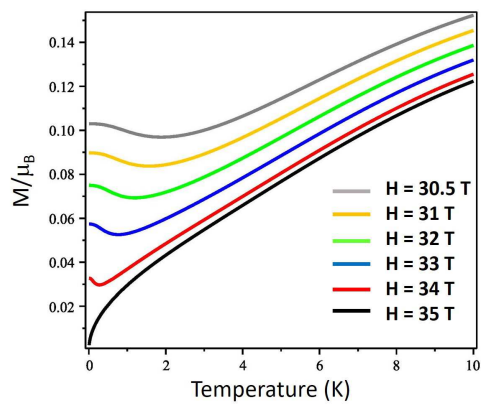}
\caption{Calculated temperature dependence of the magnetization per spin for magnetic fields $H \ge H_{c1}$ in the framework of the phenomenological model  presented here. The shown curves were computed at the magnetic fields indicated.} 
\label{fig10}
\end{figure}

\begin{figure}
\centering
\includegraphics[width=0.85\linewidth]{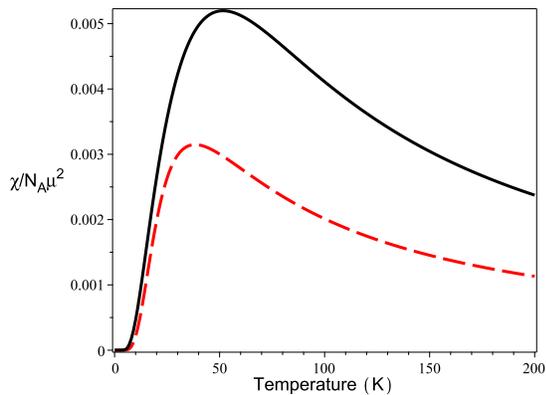}
\caption{Calculated temperature dependence of the magnetic susceptibility per spin at $H=0$ for the phenomenological model (black), and the simpler mean-field \cite{Singh_2007} (dashed red). The parameters for the latter are taken from Aczel et al.\cite{2009_aczel} $N_A$ is Avogadro's number.} 
\label{fig11}
\end{figure}

Fig.~\ref{fig10} shows the magnetization vs temperature, calculated using our phenomenological model. We see that the magnetization shows minima at $T=T_{c}(H)$, similar to the ones observed for the spin-dimer compound TlCuCl$_3$\cite{Nikuni_2000}. The dimer mean-field theory \cite{Tachiki_1970}, in contrast, predicts a constant magnetization below the critical temperature. In Fig.~\ref{fig11}, we plot the temperature dependence of the magnetic susceptibility calculated within our phenomenological theory and a simple mean-field theory result \cite{Singh_2007}. 

\begin{figure}
\centering
\includegraphics[width=0.95\linewidth]{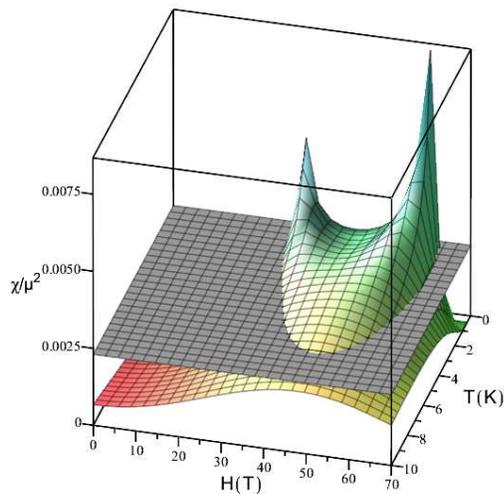}
\caption{Calculated temperature and magnetic field dependence of the magnetic susceptibility per spin. The intersection with a plane (gray) reproduces the critical ($H_c$,$T_c$) line.} 
\label{fig12}
\end{figure} 

\begin{figure}
\centering
\includegraphics[width=0.85\linewidth]{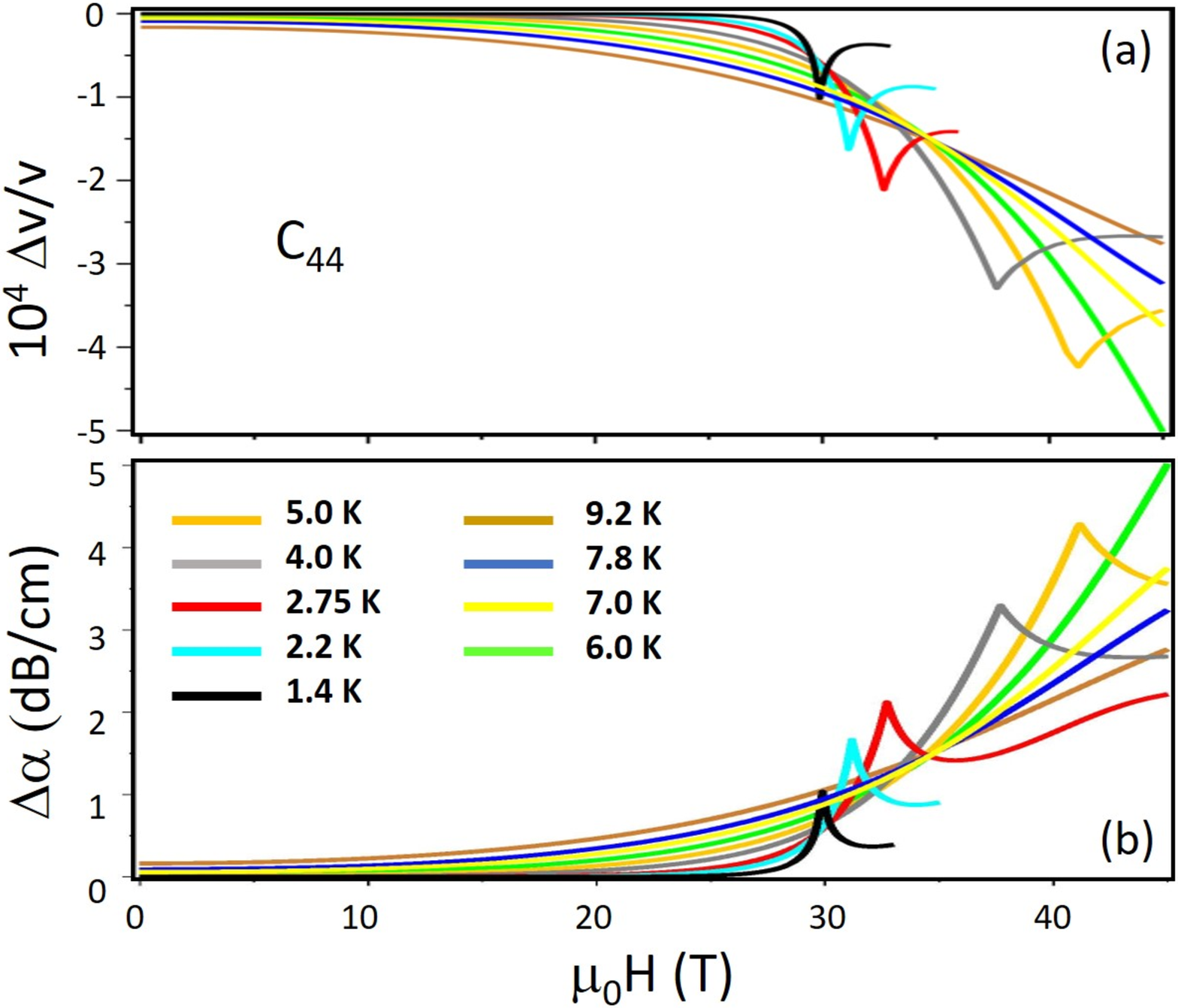}
\caption{(a) Calculated magnetic-field dependence of the relative change of the velocity of the shear acoustic mode $c_{44}$ for various temperatures indicated. (b) Calculated magnetic field dependence of the sound attenuation of the shear mode $c_{44}$ at the same temperatures. Compare to Fig.~\ref{fig6}. The gradual broadening of the transition is likely the results of higher temperatures.} 
\label{fig13}
\end{figure}

\begin{figure}
\centering
\includegraphics[width=0.85\linewidth]{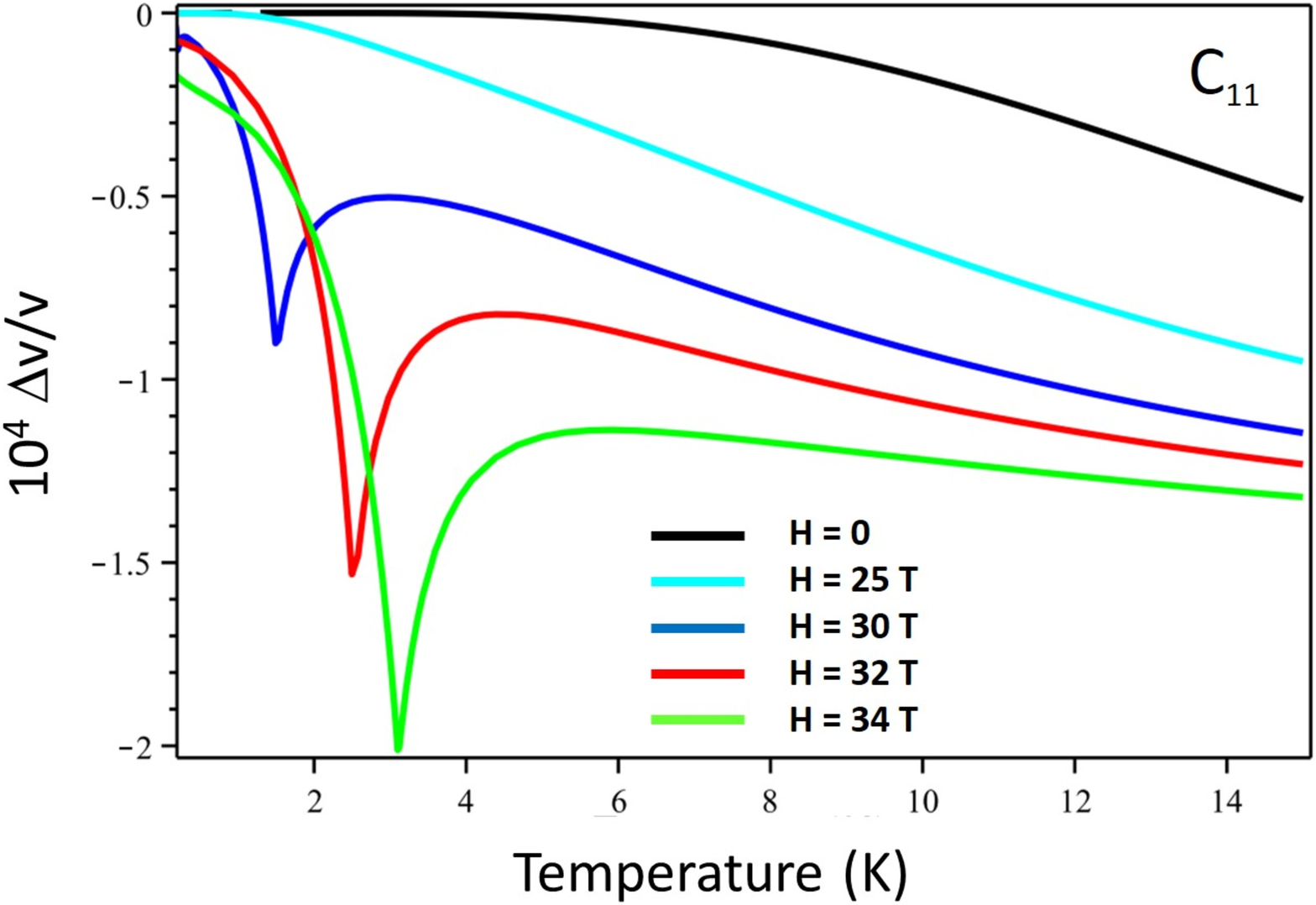}
\caption{(Color online) Calculated temperature dependence of the relative change of the velocity of the longitudinal acoustic mode $c_{11}$ for various values of the magnetic field.  Compare to Fig. \ref{fig3}(a).} 
\label{fig14}
\end{figure}

Fig.~\ref{fig12} shows the critical ($H_c$,$T_c$) line calculated in the framework of the used phenomenological model, reminiscent of the phase diagram of Sr$_3$Cr$_2$O$_8$. Note, the somewhat asymmetric nature of the dome, which is not unusual for the BEC of magnons~\cite{2014_zapf, kohama_2012}, but is absent in the mean-field dimer theory \cite{Tachiki_1970}. Now we are in a position to calculate the magneto-acoustic characteristics of Sr$_3$Cr$_2$O$_8$. The corresponding results for the acoustic properties of the $c_{44}$ shear and $c_{11}$ longitudinal modes are shown in Figs. \ref{fig13} and \ref{fig14}, respectively. There is a good agreement between the computations within the phenomenological model and our experimental observations.

\section{Summary}

Magnetocaloric effect, ultrasound, and dilatometry experiments were conducted in Sr$_3$Cr$_2$O$_8$ single-crystals that reveal coupling of spin to the crystal-lattice degrees of freedom. Our measurements of the magnetocaloric-effect in pulsed magnetic fields to 60 T show that the sample cools down dramatically when magnetized adiabatically, clearly displaying field-induced magnetic order below $T_{{\rm max}} = 8$~K as well as regions of accumulated entropy due to the spin-spin correlations for $T >T_{{\rm max}}$. We, hence, conclude that this experimental tool is an invaluable ally for studies of strongly correlated systems in the challenging environment of pulsed magnetic fields. Dilatometry experiments performed using the optical FBG method in continuous magnetic fields to 45~T were used to complete temperature sweeps at constant magnetic fields, and field sweeps at a constant temperature. While the observed crystal-lattice response at the {\it XY}-AFM phase transition  $\Delta l/l \simeq 10^{-6}$ is relatively small, the boundaries were mapped out without difficulty thanks to the high experimental resolution approaching one part in a hundred million. We find in these studies that the crystallographic $ab$-plane and $c$-axis shrink upon entering the ordered state, a behavior that is reminiscent of that previously observed in the related field-induced {\it XY}-type AFM compound NiCl$_2$-4SC(NH$_2$)$_2$. \cite{2008_zapf} Measurements of the speed of sound reveal significant softening of various ultrasound modes near the AFM (H,T) phase boundaries, that is consistent with enhanced correlations as expected for the spin-dimerized system. The observed acoustic properties are explained in the framework of a proposed phenomenological model. The reported phenomenology is  consistent with a 'passive' crystal lattice that acts as a silent witness to the mechanisms driven by magnetic correlations. While we have not identified any feature that can quantitatively be attributed to a spin liquid, the fluctuations above 8 K in magnetic fields in the range $H_{c1} < H < H_{c2}$ are a dominant effect in the MCE, the ultrasound, and the dilatometry. Further experimental studies and modeling are necessary to better understand the origin and impact of spin correlations on the results here presented.

\vspace{0.5cm}
\begin{acknowledgements}
We thank Prof. A. Loidl and Dr. Z. Wang for insightful discussions. We acknowledge the support of the HLD at HZDR, member of the European Magnetic Field Laboratory (EMFL). The research was partially supported by the DFG through SFB 1143, the W\"urzburg-Dresden Cluster of Excellence on Complexity and Topology in Quantum Matter {\it ct.qmat} (EXC 2147, project No 390858490), and by the BMBF via DAAD (project No 57457940). Work at the NHMFL was supported by the National Science Foundation through Cooperative Agreement No. DMR-1644779, the State of Florida, and the US DOE Office of Basic Energy Science project "Science at 100 T".
\end{acknowledgements}


\begin{thebibliography}{99}
\bibitem{Fermi} The Fermi point is the value of the quantum number at which the the energy of the eigenstate is zero, separating negative energy states with filling factor equal to 1 from positive energy states with filling factor equal to 0.
\end{thebibliography}
\end{document}